\documentclass[12pt,twoside,english]{article}
\usepackage[T1]{fontenc}
\usepackage[latin1]{inputenc}
\usepackage{geometry}
\usepackage{babel}
\usepackage{graphics}

\begin{document}

\begin{titlepage}

\setcounter{page}{1}
\rightline{}
\vfill
\begin{center}
 {\Large \bf Using and Constraining Nonforward Parton Distributions}\\
\vspace{.5cm}
{\bf Deeply Virtual Neutrino Scattering in Cosmic Rays\\
 and Light Dark Matter Searches
\footnote{Based on a talk presented by Claudio Corian\`o at {\bf QCD@work 2003}, Conversano, Italy, June 2003}}

\vfill
\vfill
{\large  Claudio Corian\`{o}, G. Chirilli and Marco Guzzi}

\vspace{.12in}
 {\it  Dipartimento di Fisica, Universit\`{a} di Lecce \\
 and INFN Sezione di Lecce \\ Via Arnesano 73100 Lecce, Italy}

\vspace{.075in}

\end{center}
\vfill

\begin{abstract}
We overview the construction of Nonforward Parton Distributions (NPD) 
in Deeply Virtual Compton Scattering (DVCS). 
Then we turn to the analysis of 
similar constructs in the weak sector (electroweak NPD's). 
We argue in favour of a possible use of electroweak DVCS (EWDVCS) 
as a rare process for the study of 
neutrinos in cosmic rays and for light dark matter detection in underground experiments.

\end{abstract}
\smallskip

\end{titlepage}

\setcounter{footnote}{0}

\def\beq{\begin{equation}}
\def\eeq{\end{equation}}
\def\beqn{\begin{eqnarray}}
\def\eeqn{\end{eqnarray}}
\def\ba{\begin{eqnarray}}
\def\ea{\end{eqnarray}}
\def\ie{{\it i.e.}}
\def\eg{{\it e.g.}}
\def\half{{\textstyle{1\over 2}}}
\def\nicefrac#1#2{\hbox{${#1\over #2}$}}
\def\third{{\textstyle {1\over3}}}
\def\quarter{{\textstyle {1\over4}}}
\def\m{{\tt -}}

\def\p{{\tt +}}

\def\slash#1{#1\hskip-6pt/\hskip6pt}
\def\slk{\slash{k}}
\def\GeV{\,{\rm GeV}}
\def\TeV{\,{\rm TeV}}
\def\y{\,{\rm y}}
\def\ds{\slash}
\def\l{\langle}
\def\r{\rangle}
\def\xprime{x^{\prime}}
\def\xprimetwo{x^{\prime\prime}}
\def\zprime{z^{\prime}}
\def\xprimbar{\overline{x}^\prime}
\def\xprim2bar{\overline{x}^{\prime\prime}}
\def\ptbold{\mbox{\boldmath$p$}_T}
\def\ktbold{\mbox{\boldmath$k$}_T}
\def\ktboldbar{\mbox{\boldmath$\overline{k}$}_T}
\def\beq{\begin{equation}}
\def\eeq{\end{equation}}
\def\tr{{\bf tr}}
\def\P{P^\mu}
\def\Pb{\overline{P}^\mu}
\def\BOX#1#2#3#4#5{\hskip#1mm\raisebox{#2mm}[#3mm][#4mm]{$#5$}}
\def\VBOX#1#2{\vbox{\hbox{#1}\hbox{#2}}}

\setcounter{footnote}{0}
\newcommand{\beqa}{\begin{eqnarray}}
\newcommand{\eeqa}{\end{eqnarray}}
\newcommand{\eps}{\epsilon}

\pagestyle{plain}
\setcounter{page}{1}

\section{Introduction}
\begin{figure}[t]
{\par\centering \resizebox*{6cm}{!}{\includegraphics{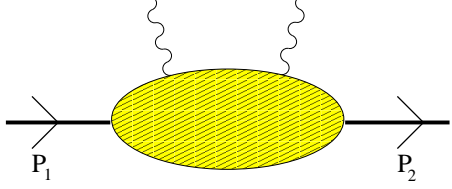}} \par}
\caption{Compton Scattering} 
\label{cs.ps}
\end{figure}

\begin{figure}[t]
{\par\centering \resizebox*{6cm}{!}{\includegraphics{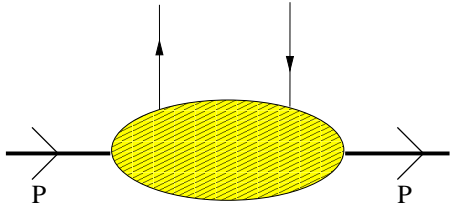}} \par}
\caption{Forward Parton Distributions} 
\label{pdf.ps}
\end{figure}

Nonforward Parton Distributions are an important construct 
in the parton model \cite{ji}, \cite{radyushkin} and appear in Compton Scattering (CS) (Fig.~\ref{cs.ps}) 
in the generalized Bjorken region \cite{muller}, where 
the process is hand-bag dominated. 
They generalize ordinary parton distributions (Fig.~\ref{pdf.ps}). 
Proofs of factorization to all orders of 
processes of this type have also been presented \cite{collins}
and there is a large amount of work (see for instance \cite{rad}) 
devoted to the subject,
accompanied by ongoing experimental efforts to measure the corresponding 
cross section \cite{sabatier}. 
In the electromagnetic case the measurements are difficult, 
given the presence of a dominant 
Bethe-Heitler background. One of the possibilities to perform the 
measurement is through the interference  between the Bethe-Heitler and the DVCS hand-bag diagram by electron spin asymmetries. 

The analysis of these distributions, including their modeling 
\cite{van}\cite{freund} and the theoretical and phenomenological study of the leading-twist 
and higher-twist contributions in the context of ordinary DVCS (with a virtual 
or a real final state photon) \cite{one} has also progressed steadily. 
In this talk we will 
briefly outline some of the features of these distributions and elaborate 
on the possibility to use some variants of them 
in the description of electroweak 
processes induced by neutral and charged currents on nucleons.
Possible applications of these generalizations which we are suggesting are in 
deeply virtual neutrino scattering for diffuse neutrinos and 
in the detection of light dark matter. Results of this analysis 
will be presented elsewhere.

\section{The DVCS Domain} 

\begin{figure}[t]
{\par\centering \resizebox*{9cm}{!}{\includegraphics{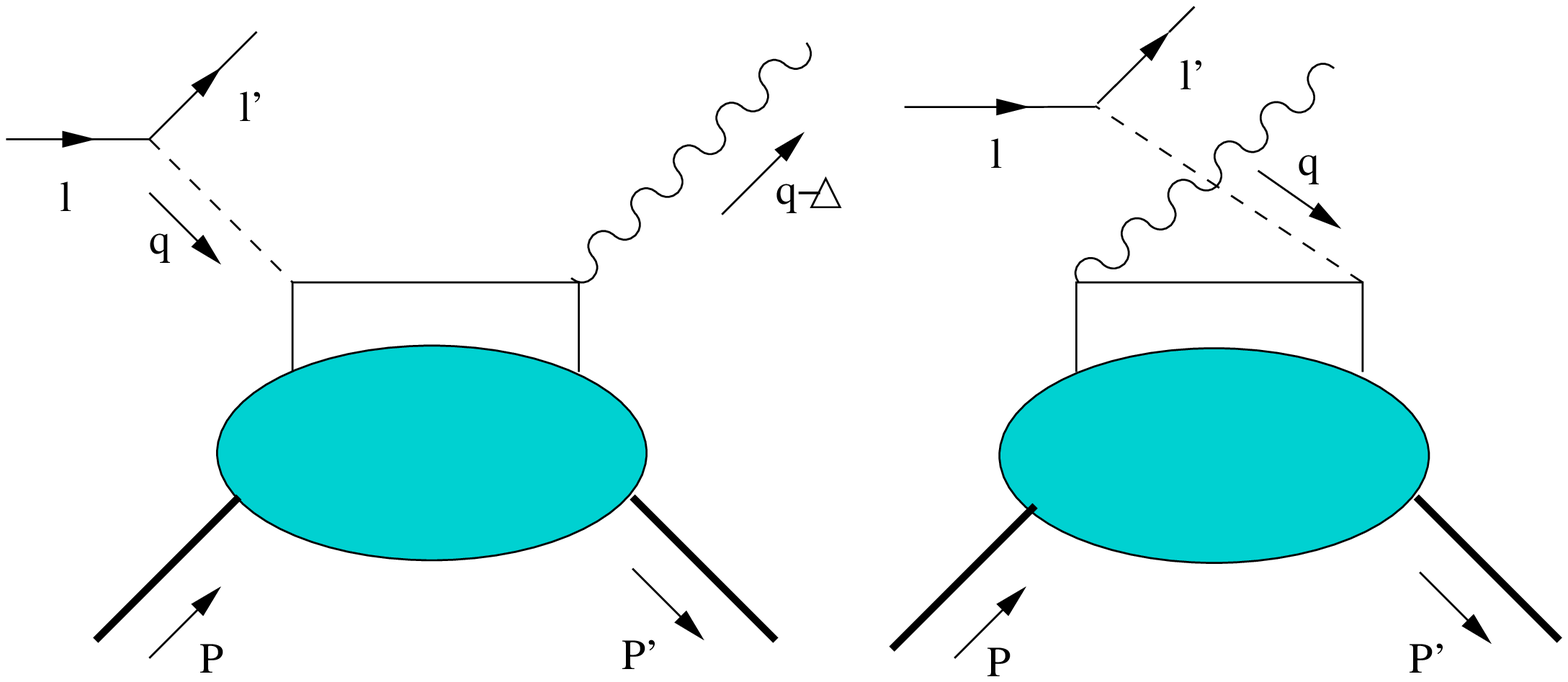}} \par}
\caption{Leading hand-bag diagrams for the process} 
\label{nfpd.ps}
\end{figure}

A pictorial description of the process we are going to illustrate is given in Fig.~\ref{nfpd.ps} where 
a lepton of momentum $l$ scatters off a nucleon 
of momentum $P_1\equiv p$ via a gauge boson exchange; 
from the final state a photon and a nucleon emerge, of momenta $q_2$ 
and $P_2\equiv=P'= p +r$ respectively. In the deeply virtual limit, 
analogous to the usual Bjorken limit, the final state photon is on shell, and  a large longitudinal (light cone) momentum exchanged is needed in order to 
guarantee factorization.   

The regime for the study of NFPD's is characterized by a deep virtuality of the 
exchanged photon in the initial interaction ($e +p\to e +p +\gamma$) ( $ Q^2 >$  2 GeV$^2$), 
with the final state photon kept on-shell, 
large energy of the hadronic system ($W^2 > 6$ GeV$^2$) 
above the resonance region and small momentum transfers $|t| < 1$ GeV$^2$. 
In the region of interest (large $Q^2$ and small $t$) 
the Bethe-Heitler background $(\sim 1/t)$ is dominant and the 
$1/Q$ behaviour of the virtual Compton scattering amplitude (VCS) 
render the analysis quite complex. 
A dedicated study of the interference BH-VCS in order to explore 
the generalized Bjorken region is therefore required.

\subsection{Nonforward Parton Distributions}  
In the case of nonforward distributions a second scaling parameter $\zeta$ 
($\xi$)
controls the asymmetry between the initial and the final nucleon momentum in the 
deeply virtual limit of nucleon Compton scattering. Both the inclusive DIS region and the exclusive ERBL region can be analized with the same 
correlator.
 We recall that in the light-cone gauge the ({\em off forward}) 
distributions $H(x,\xi)$ \cite{ji} is defined as
\beq
H_q(x,\xi,\Delta^2))= \frac{1}{2}\int \frac{dy^-}{2 \pi}e^{-i x \bar{P}^+y^-}
\langle P'| \bar{\psi}_q(0,\frac{y^-}{2},{\bf 0_\perp})
\frac{1}{2}\gamma^+ \psi_q (0,\frac{y^-}{2},{\bf 0_\perp})|P\rangle
\eeq
with $\Delta= P' - P$,$\Delta^2\equiv t$  $\bar{P}^+=1/2(P + \bar{P})$ \cite{ji} (symmetric choice) and 
$\xi \bar{P}=1/2\,\,\Delta^+$.

This distribution describes for $x>\xi$ and $x < -\xi$ the DGLAP-type region for the quark and the antiquark distribution respectively, and the ERBL 
(see \cite{rad}, \cite{coli}) distribution amplitude 
for $-\xi <x < \xi$. In the following we will omit the $\Delta$ dependence from $H_q$.

The most common procedure is to use double distributions, $F(x,y)$ defined with a symmetric choice of the external momenta  
and relate them to off-forward distributions $H(v,\xi,t)$ \cite{ji}, incorporating in a single interval with   $-1 \leq v \leq 1$ 
both the quark and antiquark parton distributions with an asymmetry parameter $\xi$, 

\beqa
&H(v,\xi,t) = \nonumber\\  
&\int^1_{-1}dx' \int^{1-|x'|}_{-1+|x'|}dy' \delta(x'+ \xi y' - v) F(x',y',t) \, ,  
\eeqa
Below we will neglect the $t$ dependence. The singlet/ nonsinglet decomposition of the evolution can be carried out as usual 
taking linear combination of flavours  
\beqa 
&H^S(v,\xi) = \sum_a H^{q,a} (v,\xi) \mp H^{q,a} (-v,\xi) \, ,\nonumber\\  
&H^{NS,a}(v,\xi) = H^{q,a} (v,\xi) \pm H^{q,a} (-v,\xi)   \, , \nonumber\\  
&H^G(v,\xi) = H^g(v,\xi) \pm H^g(-v,\xi). 
\eeqa

The evolution can be analized in various ways. One possibility is to 
use nondiagonal distributions \cite{GBM}, related to the asymmetric distributions introduced by Radyushkin (see \cite{rad})  
\beqa
&{\cal F}^{q,a} \left (X_1=\frac{v_1 + \xi}{1+\xi},\zeta\right ) = \frac{H^{q,a} (v_1,\xi)}{(1-\zeta/2)} \, , \label{fq} \\  
&{\cal F}^{{\bar q},a} \left (X_2=\frac{\xi - v_2}{1+\xi},\zeta\right ) = -\frac{H^{q,a} (v_2,\xi)}{(1-\zeta/2)} \, , \label{fqbar}  
\eeqa
where $v_1 \in [-\xi,1], \, v_2 \in [-1,\xi]$.  
The parameter $\zeta \equiv r_+/p_+ $ characterizes 
the so-called $skewedness$ of the process and is defined in the interval $ \zeta \in [0,1]$.  
Variables $\xi$ and $\zeta$  are related in the DVCS limit by $\xi = \zeta/(2-\zeta)$ and $\zeta = x_{bj}$, conciding with Bjorken variable x 
(modulo terms of  ${\cal O} (x_{bj} t / Q^2)$). 

The inverse transformations between the $v$s and $X$s are easily found in the various allowed regions 

\beqa  
v_1 = \frac{X_1 - \zeta/2}{1-\zeta/2} & \, , \, & v_2 = \frac{\zeta/2 - X_2}{1-\zeta/2} \, . 
\eeqa  

\beqa  
{\cal F}^{g} (X,\zeta) &=& \frac{H^g (v, \xi)}{(1-\zeta/2)} \, . 
\eeqa  
In the gluon case one can use any either $v_1$ or $v_2$ equivalently. 

One finds 
\beqa  
&{\cal F}^{q,a} (X,\zeta) = \frac{H^{q,a} (v_1,\xi)}{1-\zeta/2} =   
\int^{1}_{-1} dx' \int^{1-|x'|}_{-1+|x'|} dy' \delta \left( x' + \xi y' - v_1 \right)   
\frac{F^{q,a} (x',y')}{\left(1-\zeta/2\right)} \, , \nonumber \\   
&{\cal F}^{{\bar q},a} (X,\zeta) = -\frac{H^{q,a} (v_2,\xi)}{1-\zeta/2} =   
\int^{1}_{-1} dx' \int^{1-|x'|}_{-1+|x'|} dy' \delta \left( x' + \xi y' - v_2 \right)   
\frac{F^{q,a} (x',y')}{\left(1-\zeta/2\right)} \, .   
\label{fqinp}   
\eeqa
\noindent with $1> v_1 > -\xi$, $-1< v_2 < \xi$. 
The generation of initial conditions for nonforward distributions 
is usually based on factorization, as suggested originally by Radyushkin, 
starting from the double distributions ($x'-y'$ factorization). These are obtained using some profile functions 
$\pi()$ which characterizes the spreading in $y'$ 
of the momentum transfer $r$, combined  with an ordinary forward parton distributions. The latter is appropriately 
extended to the $x'<0$ to describe antiquark components, as pointed 
out long ago \cite{Jaffe}. For instance, in the quark/gluon case one obtains 
\cite{freund} \cite{radyushkin}

\beqa 
F^{q,a}(x',y') &=& \pi^{q} (x',y') f^{q,a} (x') \nonumber\\  
    &=& \frac{3}{4} \frac{(1-|x'|)^2 - {y'}^2}{(1-|x'|)^3} f^{q,a} (x') \, , \nonumber\\  
F^g(x',y')  &=& \pi^g(x',y') f^g (x') \nonumber\\  
    &=& \frac{15}{16} \frac{((1-|x'|)^2 - {y'}^2)^2}{(1-|x'|)^5} f^g (x') \, , \label{ddinp} 
\eeqa 
\noindent for quark of flavour $a$, where the quarks and gluon distributions 
are extended to $x<0$, as 
\beqa  
f^g (x)  &=& xg(x,Q_0) \Theta(x) + |x| g (|x|,Q_0) \Theta (-x) \, , \nonumber \\  
f^{q,a} (x) &=& q^{a} (x,Q_0) \Theta(x) - (\bar q^{a}) (|x|,Q_0) \Theta (-x) \, . \label{pdfinp} 
\eeqa    

\beqa   
&{\cal F}^{q,a} (X,\zeta) = \frac{2}{\zeta}  
\int^{\frac{v_1+\xi}{1+\xi}}_{\frac{v_1-\xi}{1-\xi}} dx' \pi^q \left (x', \frac{v_1 - x'}{\xi} \right) q^a (x') \, \label{dglapq}.  
\eeqa   
For the anti-quark, since $v_2 = -v_1$ one may use  
eqs.(\ref{fqbar},\ref{fqinp}) with $v_2 \to -v_1$, and,   
exploiting the fact that $f^{q} (x) = - {\bar q} (|x|) $ for $ x < 0  
$, one arrives at   
\beqa   
&{\cal F}^{\bar q,a} (X,\zeta) =  
\frac{2}{\zeta}  
\int^{\frac{-v_1+\xi}{1-\xi}}_{\frac{-v_1-\xi}{1+\xi}} dx' \pi^q \left   
(x',\frac{-v_1 - x'}{\xi} \right ){\bar q}^a(|x'|) \label{dglapbq}.   
\eeqa

The expression for the gluons is similar. Performing the y' integration, one obtains in the two regions, DGLAP and ERBL respectively 
\beqa
&{\cal F}^{g} (X,\zeta) =
\frac{2}{\zeta} \int_{\frac{X-\zeta}{1-\zeta}}^{X} dx' \pi^g \left(x', {\tilde y}(x') \right ) ~x'g(x') \, , 
\eeqa

and
\beqa
{\cal F}^{g} (X,\zeta) &=
\frac{2}{\zeta}\Big[ \int_{0}^{X} dx' \pi^g \left(x', {\tilde y}(x') \right ) ~x'g(x')\nonumber\\
&+ \int_{0}^{\zeta-X} dx' \pi^g \left(x', {\tilde y}(-x') \right ) ~x'g(x')\Big]. 
\eeqa
\subsection{A Comment}
\begin{figure}[t]
{\par\centering \resizebox*{6cm}{!}{\includegraphics{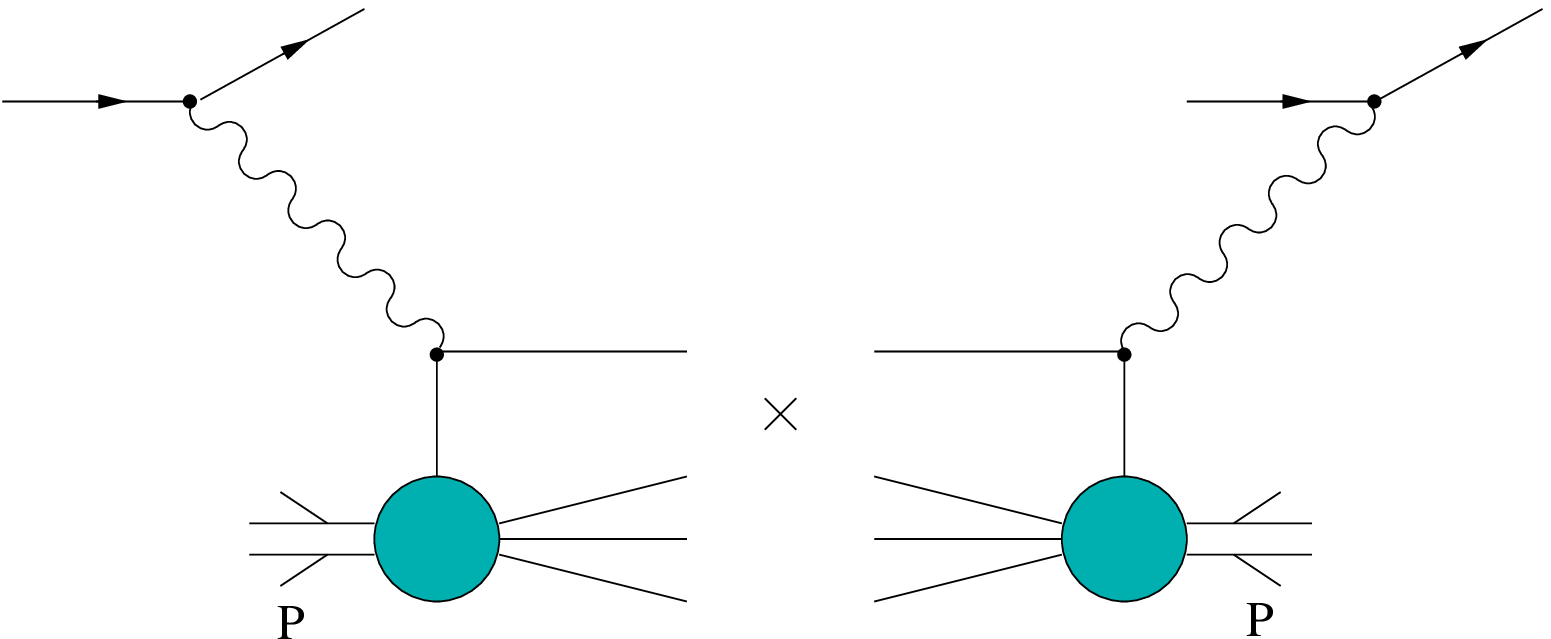}} \par}
{\par\centering \resizebox*{6cm}{!}{\includegraphics{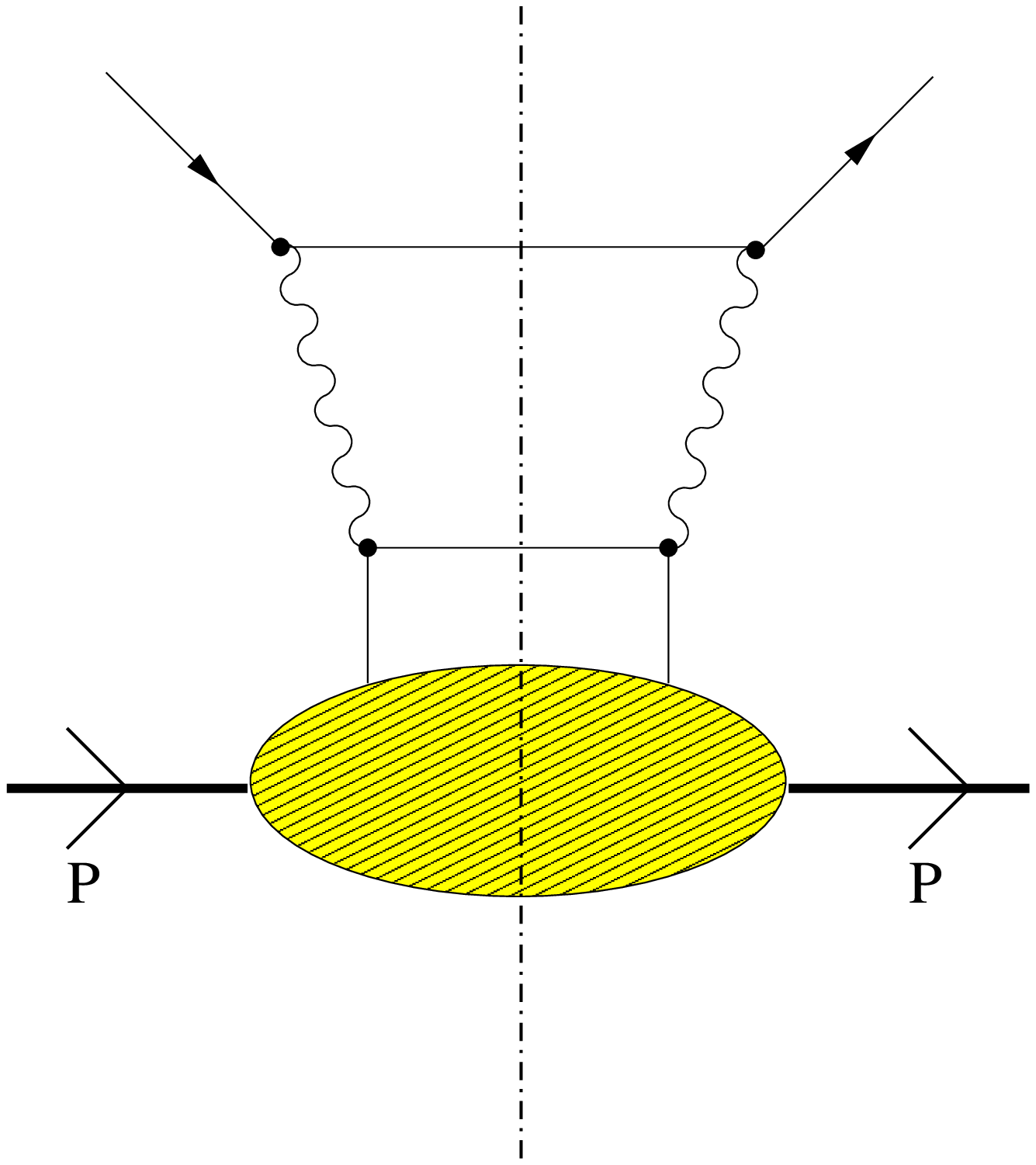}} \par}
\caption{Unitarity diagram in DIS (forward Compton scattering)} 
\label{uni.ps}
\end{figure}

Parton distributions are quasi-probabilities, very similar 
to Wigner functions (as first observed in \cite{ccg}), 
though the momentum dependence of the latter 
is not found easily found in the former. One possibility in this direction 
has been suggested recently \cite{ji1}.
The formal definition of a 
parton distributions is in a non-local light cone correlator, built 
through unitarity (at least in the forward case). In the 
nonforward case there is no optical theorem (Fig.~\ref{uni.ps}) that holds, and the construct is 
a genuine correlation function in the nucleon state. 
We have pointed out in \cite{ccg} the existence of a formal 
relation between evolution equations and kinetic equations, via a Kramers-Moyal expansion of the non-forward evolution equation. 
This relation carries a strong similarity to the well known 
relation between Wigner functions and their associated 
evolution equations, given by differential operators of arbitrarily high 
orders (Moyal products) as known from the phase-space approach 
to quantum mechanics \cite{cosmas}. 
Indeed the definition of a parton distribution takes place through 
a Wigner-Weyl transform, limited to the light cone domain.  

The evolution equations describing NFPD's are known in operatorial form 
\cite{bali}. 
Single and double parton distributions are obtained sandwiching the operatorial solution with 4 possible types of initial/final states $<p|...|p>, <p|...|0>, <p'|...|p>$, corresponding, respectively, 
to the case of diagonal parton distributions, distribution amplitudes and, in the latter 
case, skewed and double parton distributions \cite{ji,bali}. 
The ansatz for the general solution of the evolution equations up to 
next-to-leading order  for $H_q(x,\xi)$ is given by \cite{ccg}
\beq 
H_q(x,\xi)=\sum_{k=0}^{\infty} \frac{A_k(x,\xi)}{k!}\log^k
\left(\frac{\alpha(Q^2)}{\alpha(Q_0^2)}\right) +
\alpha(Q^2)\sum_{k=0}^\infty \frac{B_k(x,\xi)}{k!}\log^k
\left(\frac{\alpha(Q^2)}{\alpha(Q_0^2)}\right),
\label{expansionx}
 \eeq
where we have introduced arbitrary scaling coefficients $(A_k, B_k)$, now functions of two parameters $(x,\xi)$.  
 
The evolution can be written down in various forms. The 
one we have exploited comes from kinetic arguments on which we now briefly elaborate. 

In the non-forward case (DGLAP-type evolution) 
the identification of a transition probability 
for the random walk \cite{ccg} associated with the evolution of the parton distribution is obtained 
via the non-forward transition probability 
\beqa
w_\zeta(x|y) &=&\frac{\alpha}{\pi}C_F \frac{1}{y- x}\left[1 + \frac{x}{y}\frac{(x-\zeta)}{y - \zeta}\right]
\theta(y>x)\nonumber \\
w'_\zeta(y|x)&=&\frac{\alpha}{\pi}C_F \frac{x^2 + y^2}{x^2(x - y)}\theta(y<x)
\eeqa
and the corresponding master equation is given by 
\beq
\frac{d \mathcal{F}_q}{d\tau}=\int_x^1 dy\, w_\zeta(x|y)\mathcal{F}_q(y,\zeta,\tau)-
\int_0^x dy\, w'_\zeta(y|x)\mathcal{F}_q(x,\zeta,\tau),
\eeq
that can be re-expressed in a form which is a simple generalization of the formula for the 
forward evolution \cite{ccg}

\begin{figure}[t]
{\par\centering \resizebox*{7cm}{!}{\includegraphics{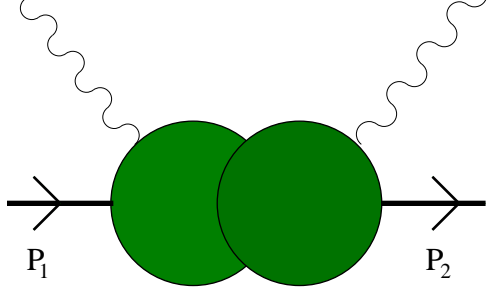}} \par}
\caption{Overlap of wave functions in proton Compton scattering} 
\label{overlap}
\end{figure}

A Kramers-Moyal expansion of the equation allows to generate a differential equation 
of infinite order with a parametric dependence on $\zeta$ 

\beqa
\frac{d \mathcal{F}_q}{d\log Q^2} &=&\int_{\alpha(x)}^{0}dy\,  
w_\zeta(x+y|x)\mathcal{F}_q(x,\zeta,\tau) + 
\int_{0}^{-x}dy\,  
w'_\zeta(x+y|x)\mathcal{F}_q(x,\zeta,\tau) \nonumber \\
&& - \sum_{n=1}^{\infty}\int_0^{\alpha(x)}dy \frac{(-y)^n}{n!}{\partial_x}^n
\left(w_\zeta(x+y|x)\mathcal{F}_q(x,\zeta,\tau)\right).
\label{expans}
\eeqa

If we arrest the expansion at the first two terms $(n=1,2)$ we are able to derive an 
approximate equation describing the dynamics of partons for non-diagonal transitions which is similar to constrained random walks \cite{ccg}. 

\begin{figure}[t]
{\par\centering \resizebox*{7cm}{!}{\includegraphics{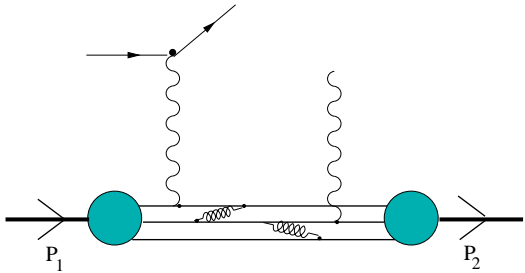}} \par}
\caption{ERBL factorization of proton Compton scattering} 
\label{erbl}
\end{figure}

\subsection{Possible Constraints}
Kinetic analogies are an interesting way 
to test positivity of the evolution for 
suitable (positive) boundary conditions, though their practical 
use appears to be modest, since it is limited to leading order 
in the strong coupling $\alpha_s$ \cite{ccg}.  
Other interesting constraints emerge also from other studies, 
based on a unitarity analysis \cite{pire}, which also may be of help. 
Their practical impact in the modeling of these distributions 
has not been studied at all. 

\begin{figure}[t]
{\par\centering \resizebox*{5cm}{!}{\includegraphics{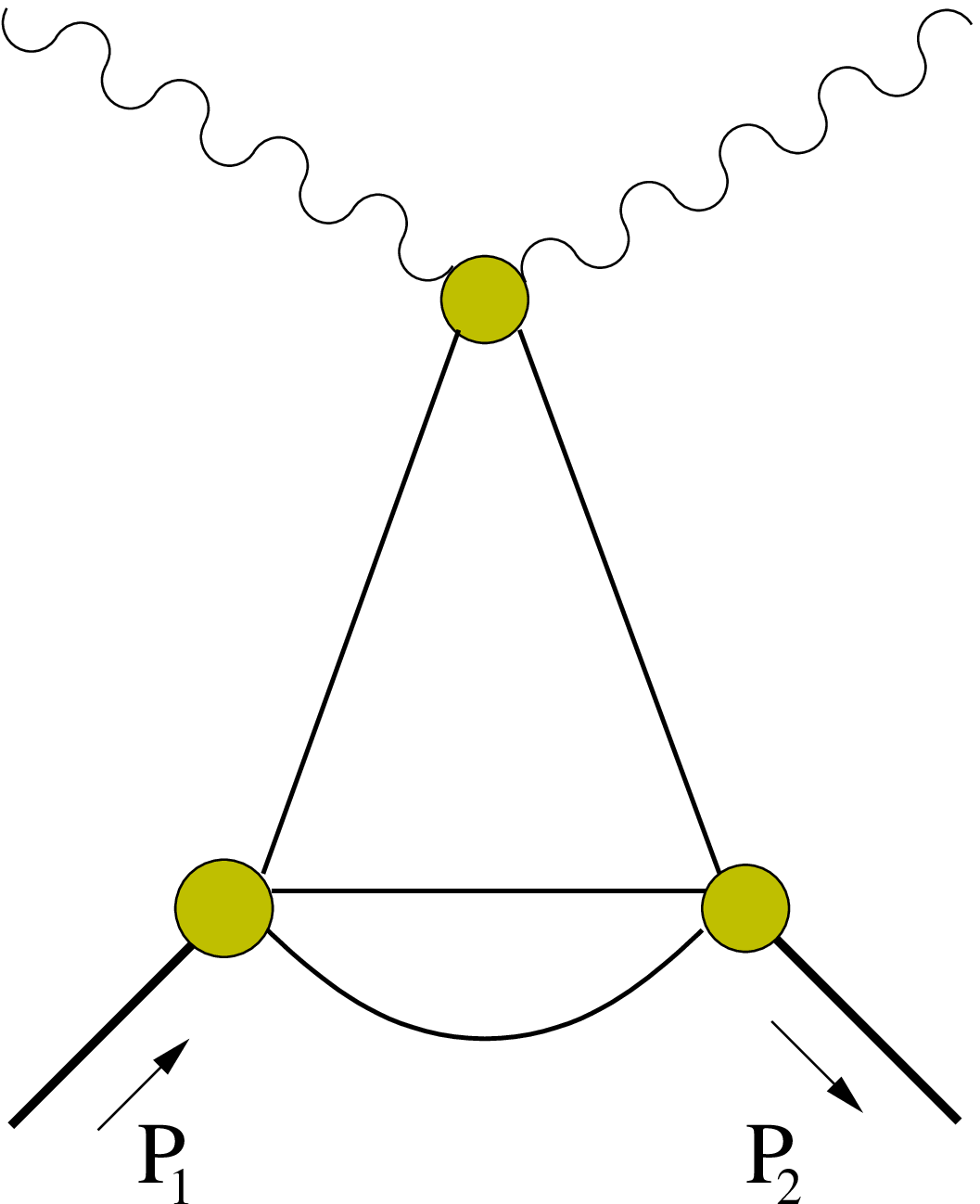}} \par}
{\par\centering \resizebox*{5cm}{!}{\includegraphics{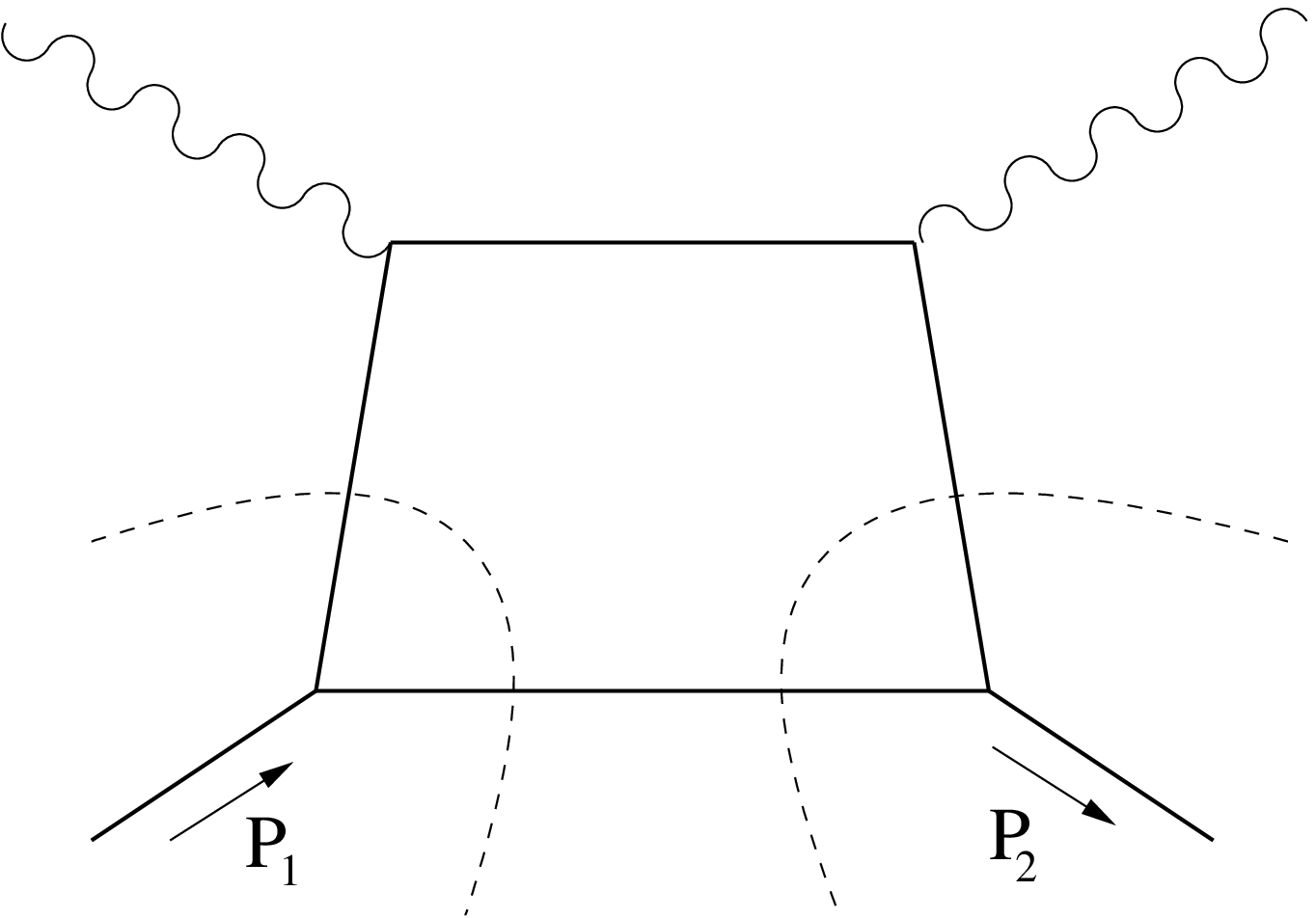}} \par}
\caption{Dispersive Approach to proton Compton scattering 
using current correlators (above). The spectral density of pion Compton scattering (below).} 
\label{dispersive}
\end{figure}

We should stress that these QCD bounds, similar to the Soffer bound for $h_1$ 
\cite{soffer},
some of them involving forward and nonforward distributions at the same time, 
should be analized carefully in perturbation theory. 
All the anomalous dimensions are known up to next-to-leading order and the study is possible. Another input comes from Ji's sum rule on angular momentum 
\cite{ji}, which needs to be satisfied. One can view Ji's sum rule also 
as a constraint between the evolution of angular momentum and the nonforward parton distributions for a vanishing $t$.
The evolution of all the distributions, including angular momentum 
is also known, and there is wide room to play with the boundary 
conditions. Other constraints may come from a combined sum rule analysis for 
proton Compton scattering, along the lines suggested in \cite{ster}. 
The soft mechanism based on the overlap of wave functions (see Fig.~\ref{overlap}), pre-asymptotic (Fig.~\ref{erbl}) may be studied using 
dispersion theory of 4-current correlators (Fig.~\ref{dispersive}), especially for scattering at wide angle.

\section{EWDVCS} 
We now come to discuss an interesting generalization of the standard DVCS process. 

The electroweak version of the process (EWDVCS),
in the regime that we are interested in, 
requires the notion of electroweak NPD's to be conveniently described. A charged current version of this interaction, with an electron 
or a neutrino in the final state, the latter coming from an electron, can also be 
contemplated. We are going to argue 
that there are motivations for studying these extensions for astroparticle applications.  

If we entertain the possibility of extending DVCS to the weak sector 
we will be facing the issue of large suppression in the cross section. 
For charged currents there is a similar BH background, but not for neutral currents, for instance 
in neutrino-nucleon scattering, this aspect is obviously not present.  
We shoukd keep in mind that it is very likely that most of our hopes to detect dark matter using ground or underground 
based facilities, whatever they may be, have to rest considerably 
on our capability to detect weakly interacting particles of various masses and 
in various kinematical domains. 

Differently from standard accelerator searches,  
in the astroparticle case the range of these interactions vary 
from low to high to extremely high energy. 

The example of Ultra High Energy Cosmic Rays (UHECR) 
\cite{ccf} and the undeterminations associated to 
the estimates of their energy 
both in the strong sector (for instance at the Greisen-Ztsepin-Kuzmin (GZK) cutoff and around it for p-p first 
impacts) and in the weak 
sector - such as in the detection of neutrino-induced horizontal air showers - are clear 
examples of a theory that is unable to be fully predictive. 
We remark that arguments linked 
to parton saturations and to the small-x impact of the distributions for 
Ultra High Energy (UHE) neutrino scattering have been formulated only recently and show 
a considerable readjustment \cite{kwiecinski} of the UHE cross section. 
In particular, 
our overall understanding of this domain is still under debate, especially 
from the S-matrix perspective \cite{alan} 
Neutrino studies have only been 
confined to the atmospheric and solar case, for obvious reasons. 
Though the fluxes may change considerably with a varying energy, 
it is conceiveable that other cosmic ray experiments may 
be looking for bounds on the parameter space of the various 
models which have been recently proposed.

In the search for dark matter it is essential to have an open view 
on the possible mass range of the possible candidates, 
including lower mass candidates \cite{celine}. 
While dark matter searches can be rather indirect and can rely on new  
astronomical searches with excellent perspectives for the future - such as weak gravitational lensing (gravitational shear measurements) or other observational methods having a statistical imprint - \cite{saas-fee}
weak or superweak interactions are still 
likely to be studied extensively using nucleon targets. We believe that 
the few GeV region of QCD has very good chances to be relevant in the 
astroparticle territory, and for a just cause. One additional photon in the final state plus a recoiling nucleon is a good signal for detecting 
light dark matter and both the fluxes and the cross sections should be accurately quantified \cite{amore}.

\section{Conclusions} 
We have illustrated some of the basic features of NFPD's and we have 
emphasized that there are other possible applications of 
the theory, especially in the weak sector, which may be relevant 
in the astroparticle domain. 
Any attempt to obtain a complete coverage of 
neutrinos and dark matter interaction with nucleons 
in underground experiments should also include searches 
in the intermediate energy region of QCD with far reaching consequences.

\centerline{\bf Acknowledgements} 
C.C. thanks Paolo Amore for collaborating (long ago) on the analysis 
of the process and Celine Boehm and A. Faraggi for discussions on dark matter searches during a visit at Oxford University, England.

\end{document}